\documentclass[%
,preprint%
,secnumarabic%
,amssymb, amsmath,nobibnotes, aps, prl]{revtex4}

\usepackage{graphicx}

\begin{document}

\title{No anomalous scaling in electrostatic calibrations for Casimir force measurements}

\author{S.~de Man}
\author{K.~Heeck}
\author{D.~Iannuzzi}
\email{iannuzzi@few.vu.nl} \affiliation{Department of Physics and Astronomy, VU University
Amsterdam, De Boelelaan 1081, 1081 HV Amsterdam, The Netherlands}

\date{\today}

\begin{abstract}
In a recent paper (Phys.~Rev.~\textbf{A78}, 020101(R) (2008)), Kim et al.~have reported a large
anomaly in the scaling law of the electrostatic interaction between a sphere and a plate, which was
observed during the calibration of their Casimir force set-up. Here we experimentally demonstrate
that this behavior is not universal. Electrostatic calibrations obtained with our set-up follow the
scaling law expected from elementary electrostatic arguments, even when the electrostatic voltage
that one must apply to minimize the force (typically ascribed to contact potentials) depends on the
separation between the surfaces.
\end{abstract}

\maketitle

Casimir force experiments are routinely used to set new limits on Yukawa corrections to the
Newtonian gravitational attraction between surfaces at submicron separation (see \cite{decca} and
references therein). To strengthen those constraints, new experiments must provide data with errors
smaller than any other previous measurement. With claims reaching 0.19\% relative experimental
errors at a 95\% confidence level \cite{decca}, it is important to ask whether there exists any
technical challenge or physical mechanism that might impede any further improvement in this
important field.

In a recent paper \cite{kim}, Kim et al. have reported a systematic effect observed during the
calibration of their Casimir force set-up that might represent a severe limitation to the
development of future experiments on the Casimir effect, and that, in some extent, might have been
overlooked even in previously reported high accuracy measurements. They observed that the
electrostatic force gradient between a $\simeq 30$ mm radius spherical mirror and a metallic plate
scales with surface separation $d$ like $\simeq 1 / d^{1.7}$, which represents a $15 \%$ deviation
on the exponent with respect to the $1 / d^2$ behavior expected from elementary electrostatic
calculations. If this anomaly were confirmed to be a general phenomenon for metallic surfaces at
very close separations, all the arguments used to calibrate high accuracy Casimir force set-ups,
which relies on elementary analysis of the electrostatic attraction between the two surfaces, would
be invalidated \cite{fluids}, with severe consequences on the results of those experiments.
Driven by these considerations, we have performed a high precision experiment to investigate the
electrostatic force between a sphere and a plate in the separation range from $\simeq 100$~nm to
$\simeq 2$ $\mu$m.

The main goal of this paper is to test the validity of:

\begin{equation}
F=-\frac{\varepsilon_0 \pi R \left(V+V_0\right)^2}{d}, \label{elecforce}
\end{equation}

\noindent where $F$ is the electrostatic force expected between a spherical surface of radius $R$
and a plate kept at a separation $d$ with $d<<R$, and where $\varepsilon_0$ is the permittivity of
vacuum, $V$ is the applied voltage, and $-V_0$ represents the voltage that one has to apply to
obtain minimal electrostatic force (typically ascribed to contact potentials)
\cite{lamoreaux_1997}.

\begin{figure}[h!]
\includegraphics[width=13cm,angle=-90]{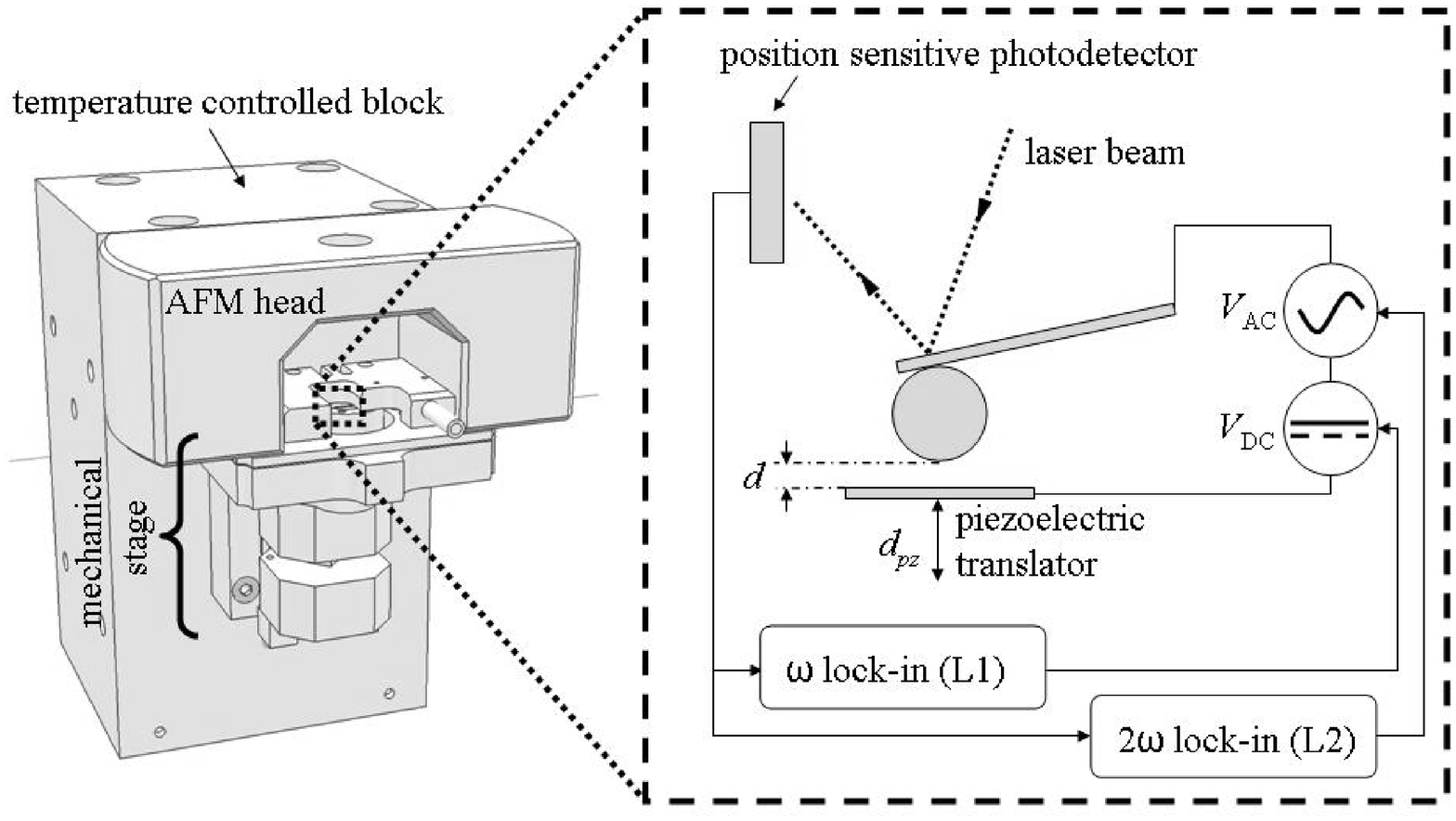}
\caption{Schematic view of the experimental apparatus.}
\label{schematic}
\end{figure}

In the apparatus used for this experiment (see Fig.~\ref{schematic}), the sphere is directly glued
under the hanging end of a micromachined cantilever (the force sensor), which is then mounted
inside the measuring head of a commercial atomic force microscope (AFM). The plate is anchored to a
custom-designed mechanical stage, which is fixed underneath the AFM head. The mechanical stage
allows one to bring the plate in close proximity with the sphere, and to perform measurements of
the force as a function of separation.

The force sensor used in this experiment is a rectangular Si cantilever ($525 \times 35 \times 4$
$\mu$m$^3$) with nominal spring constant $k \simeq 0.9$ N/m. A 100~$\mu$m radius polystyrene
divinylbenzene sphere (Duke Scientific) is attached to the free end of the cantilever with UV
curable glue. The cantilever and the sphere are then coated with a 10~nm Ti adhesion layer and a
200~nm Au film by magnetron sputtering. A similar coating is also deposited onto the plate -- a 5
$\times$ 2~mm$^2$ polished sapphire slide. The AFM head is a low-noise Veeco Multimode, which
exploits standard optical lever techniques to measure deflections of the cantilever with 0.1~nm
precision over a 50~kHz bandwidth. The mechanical stage consists of a stick-slip motor (Attocube)
and a piezoelectric translator (PI). The stick-slip motor is moved only at the beginning of the
experiment to bring the plate within a few microns from the sphere, while the separation between
the two surfaces during the actual experiment is varied with the piezoelectric translator. This
translator, which is controlled by a capacitive feedback loop, has been calibrated by the
manufacturer against traceable standards. Its resolution is reported to be equal to 50~pm. The
mechanical stage and the AFM head are anchored to a 1~$\mathrm{dm}^3$ aluminum block that is
maintained at a temperature $\simeq 10$~K above room temperature by means of a feedback controlled
heating system. The block is mounted on a commercial active anti-vibration stage (Halcyonics),
which is placed inside a 1~$\mathrm{m}^3$ acoustic isolation box. Finally, this box is placed on an
optical table inside a temperature controlled room.

In order to measure the dependence of the electrostatic force between the sphere and the plate as a
function of their separation, we apply an oscillating voltage $V=V_{DC} + V_{AC} \sin \left( \omega
t \right)$. Using Hooke's law and Eq.~\ref{elecforce}, the signal $S$ of the photodetector of the
AFM head can be written as

\begin{equation}
S=\gamma \frac{F}{k}=-\frac{\gamma \varepsilon_0 \pi R}{k d^p}
\left[\left(V_0+V_{DC}\right)^2+\frac{V_{AC}^2}{2} +2\left(V_0+V_{DC}\right) V_{AC} \sin\left(
\omega t \right)-\frac{V_{AC}^2 \cos\left(2 \omega t\right)}{2}\right], \label{elecforceharmonics}
\end{equation}

\noindent where $k$ is the spring constant of the cantilever and $\gamma$ is a parameter that
translates cantilever deflection into signal (expressed in V/m). Note that we have added an
exponent $p$ to the denominator: our goal is to verify whether $p=1$ or not. The signal $S$
contains two static components and two periodic components at angular frequencies $\omega$ and $2
\omega$. The static components of the signal represent the static deflection of the cantilever,
which, as we will show later, is always smaller than 0.2 nm, and will thus be neglected. The two
periodic components can be used to fully characterize the electrostatic interaction. For this
reason, the photodetector is connected to two lock-in amplifiers operating at frequencies $\omega$
(L1) and $2\omega$ (L2), as illustrated in Fig.~\ref{schematic}. From Eq. \ref{elecforceharmonics},
one can see that the output of L1 is proportional to $V_0+V_{DC}$. One can thus create a negative
feedback loop where L1 generates $V_{DC}$ in such a way that it keeps $V_0+V_{DC}$ small, as
typically done in Kelvin probe force microscopy \cite{kelvinprobe}. The loop gain $G$ of the
current experiment varies from $10^3$ (at $\simeq 2 \mu$m separation) to $10^4$ (at $\simeq 100$nm
separation). Because $V_{DC}=-\frac{G}{G+1}V_0$ and $|V_{DC}| < 50$mV in all our measurements, the
feedback loop certainly compensates $V_0$ down to $|V_{DC}+V_0| < 50~\mu$V. Thus, one can assume
$V_{DC}=-V_0$ for all practical purposes. We stress that the purpose of this feedback loop is
two-fold: the compensation voltage $V_{DC}$ is measured accurately at all distances and the static
deflection of the cantilever is minimized by effectively zeroing the first term of the expansion in
Eq.~\ref{elecforceharmonics}. As far as L2 is concerned, note that its peak-to-peak value is given
by:

\begin{equation}
S_{2\omega}=\frac{\gamma \varepsilon_0 \pi R}{k d^p}V_{AC}^2 \equiv \alpha V_{AC}^2,
\label{alphadefinition}
\end{equation}

\noindent where $\alpha$ is proportional to the curvature of the parabola described by
Eq.~\ref{elecforce}, which can be obtained as $S_{2 \omega} / V_{AC}^2$. Therefore, by examining
the measured values of $\alpha$ as a function of $d$ one can verify whether $p=1$ or not. To obtain
$\alpha$ as a function of $d$, we start by placing the plate a few micron away from the sphere. We
then move the plate towards the sphere in discrete steps. For each position, we measure
$S_{2\omega}$ for a properly chosen value of $V_{AC}$. At first, one might think to simply keep
$V_{AC}$ constant during the whole run. However, it is more convenient to reduce $V_{AC}$ such that
$S_{2\omega}$ stays constant as the surfaces approach, because, in this way, the relative random
error of $\alpha$ is equal for all separations: $\sigma_\alpha / \alpha = \sigma_{S_{2\omega}} /
S_{2\omega}$ (if we assume that the random error on $V_{AC}$ is negligibly small). Therefore,
before we move the plate to the next measurement point, we use the values of $\alpha$ measured in
the same run at larger separations to estimate the value that $\alpha$ should assume in the next
position, and we set the new value of $V_{AC}$ accordingly \cite{only8datapoints}. This procedure
does not by any means introduce systematic errors on $\alpha$, which is still evaluated as
$\alpha=S_{2 \omega} / V_{AC}^2$, where $S_{2\omega}$ is the actual value measured by L2.
Systematic errors on $V_{AC}$ can also be ruled out, because we calibrated the digitally controlled
function generator before starting the experiment.

In the current experiment, L2 has a 24dB/octave low pass filter with a 1 s RC time, and data are
acquired with 5 s integration time at every position. We use an $S_{2 \omega}$ set-point that
corresponds to a peak-to-peak cantilever movement at $2\omega$ of roughly 0.3~nm
\cite{determine_gamma}. With this set-point, $V_{AC}$ varies between 450~mV at $d\simeq$~2~$\mu$m
to 100~mV at $d\simeq$~100~nm. This corresponds to a static deflection of the cantilever at
$V_{DC}=-V_0$ of roughly 0.15 nm, which is thus negligible. To minimize drifts of the amplitude
response $S_{2 \omega}$ due to potential changes of the resonance frequency of the cantilever, we
work in the quasi-static regime, setting $\omega / 2 \pi$ to 72.2 Hz \cite{delyiannis-friend},
which is much smaller than the resonance frequency of the force sensor (1.65 kHz, as obtained with
an independent measurement). The total measurement consists of 184 runs over 1050 minutes.

\begin{figure}[h!]
\includegraphics[width=13cm]{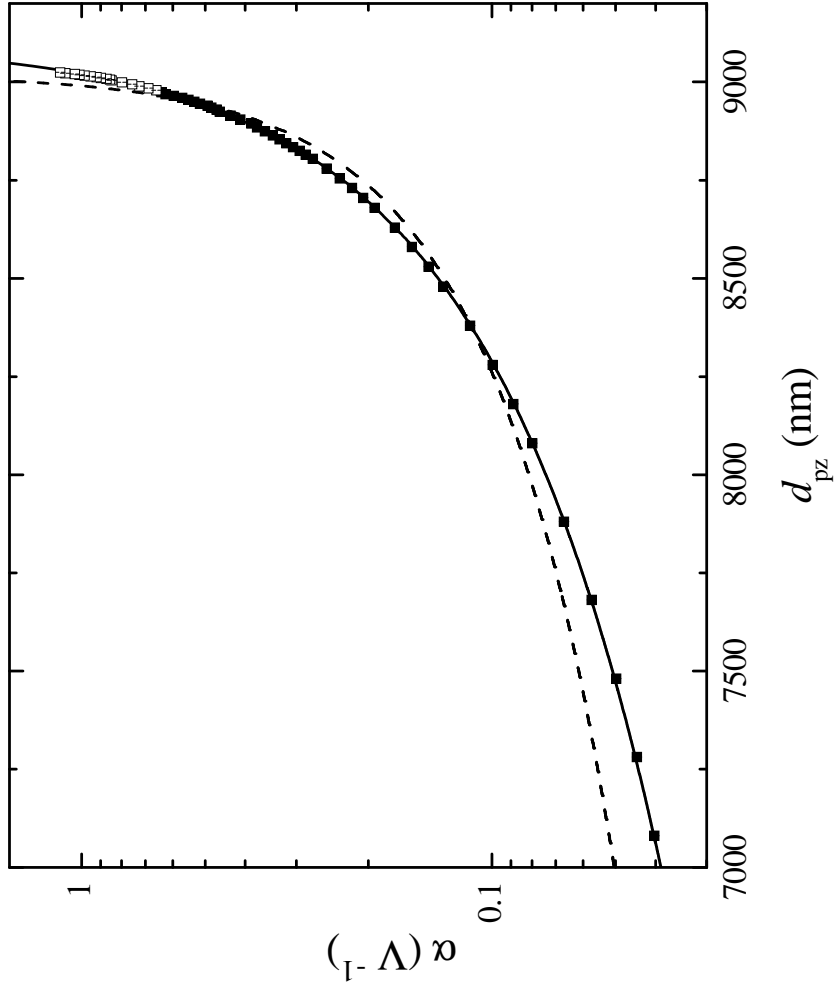}
\caption{Plot of $\alpha$ (see Eq.~\ref{alphadefinition}) as a function of the position of the
piezoelectric translator for run number 107. The error bars on the data are within the dimension of
the symbol. Black squares indicate data that are used for the analysis. Open squares are data that
are excluded from the analysis. The continuous line shows the fitting curve obtained with
elementary electrostatic arguments ($\alpha \propto 1/d$). The dashed line represents the best fit
obtained on the basis of the anomalous behavior observed in \cite{kim} ($\alpha \propto
1/d^{0.7}$).} \label{run107}
\end{figure}

\begin{figure}[h!]
\includegraphics[width=13cm]{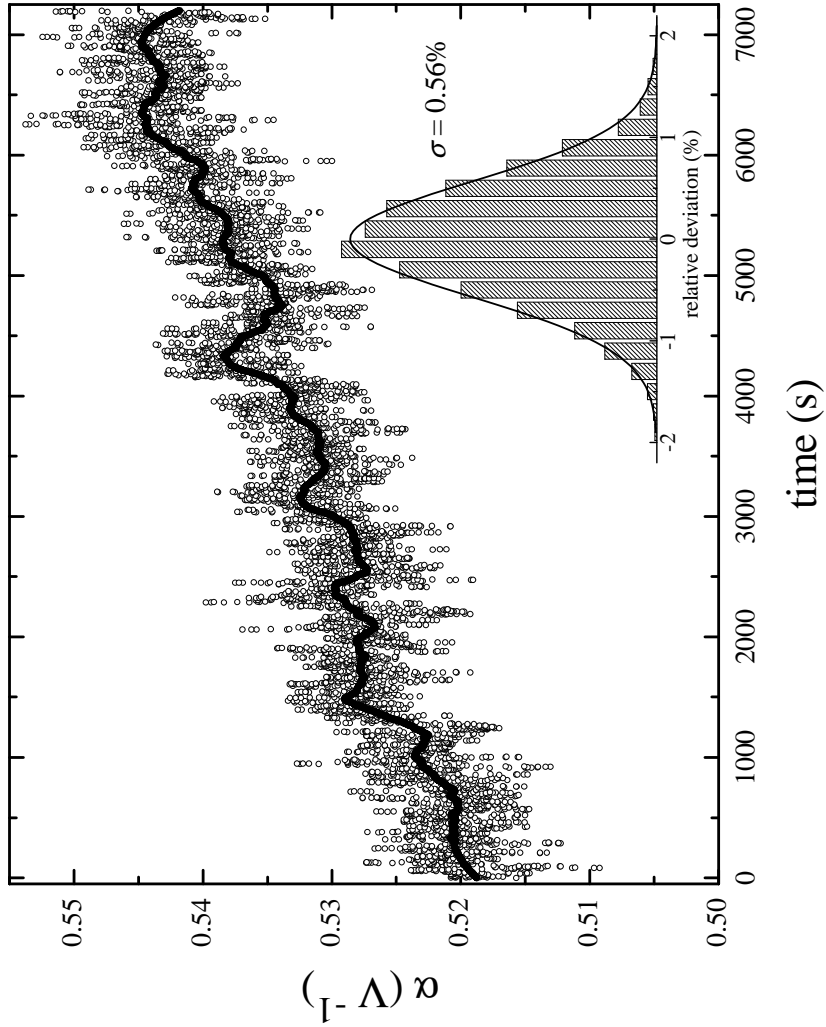}
\caption{Plot of $\alpha$ (see Eq.~\ref{alphadefinition}) as a function of time for a fixed value
of $d_{pz}$. The thick line represents a smoothed curve that accounts for the drifts in our set-up
during this measurement. The inset shows a histogram of the relative difference between the data
points and the smoothed curve, together with a Gaussian fit.} \label{alphaerr}
\end{figure}

In Fig.~\ref{run107} we show the value of $\alpha$ as a function of the extension of the
piezoelectric stage $d_{pz}$ (see Fig.~\ref{schematic}) for one randomly chosen run. If we neglect
the static deflection of the cantilever, the actual separation between the surfaces $d$ is given by
$d=d_0-d_{pz}$, where $d_0$ is the initial separation, which is \emph{a priori} unknown. To
validate whether $p=1$, we fit the data with an equation of the form

\begin{equation}
\alpha = \frac{\kappa}{\left(d_0-d_{pz}\right)^p} \label{alphafit},
\end{equation}

\noindent where $d_0$ and $\kappa=\frac{\gamma \varepsilon_0 \pi R}{k}$ are free parameters. The
fits are performed using standard $\chi ^2$ minimization algorithms, for which it is necessary to
first estimate the error on the data. This is done by measuring $\alpha$ at a fixed $d_{pz}$ for
120 minutes. The results are shown in Fig.~\ref{alphaerr}. The data distribute along a smooth curve
that is not constant because of drifts in $d_0$ and/or $\kappa$ \cite{noteondrifts}. In the inset,
we plot a histogram of the relative difference between the data points and the smoothed curve. One
can clearly see that $\alpha$ follows a normal distribution with a standard deviation of 0.56\%;
the relative error in a single measurement of $\alpha$ is thus 0.56\%. Note that this error
represents an uncertainty of 600~fm in the determination of the root-mean-square motion in response
to the varying potential difference $V_{AC}$ \cite{determine_gamma}. Using this relative error, we
repeat the fit three times: letting $p$ as a free parameter, and forcing $p=1$ (as elementary
electrostatic calculations suggest) or $p=0.7$ (as found in \cite{kim}). If $p$ is a free
parameter, one obtains $p=1.005 \pm 0.004$ with reduced $\chi ^2$ equal to 1.19. The fit with $p=1$
gives rise to a comparable value of reduced $\chi ^2$ ($\chi ^2=1.21$). The fit with $p=0.7$, on
the contrary, produces a reduced $\chi ^2$ of 411. Our data thus follow the behavior expected from
elementary electrostatic arguments. It is important to stress that these values are obtained
analyzing all data except those for which $d<120$ nm (see the open squares of Fig.~\ref{run107}).
If those data are included, the $\chi ^2$ quickly increases. This is not surprising because, at
small separations (smaller than $\simeq 120$ nm), the Casimir force bends the cantilever so much
that, within the precision of the current experiment, one cannot assume $d=d_0-d_{pz}$ anymore.

To make our claim more robust, we fit each single data set with $p=1$ and analyze the behavior of
the reduced $\chi ^2$. For the sake of computational convenience, we rewrite Eq. \ref{alphafit} as

\begin{equation}
\frac{1}{\alpha}=\frac{1}{\kappa} \left(d_0-d_{pz}\right). \label{1overalphaequation}
\end{equation}

\noindent To avoid systematic errors due to the bending of the cantilever at small separations, we
apply again a mask to the closest $N-41$ data points, where $N$ is the total number of points in a
single run (similar to Fig.~\ref{run107}). The relative error on $\alpha$ is so small, that
$1/\alpha$ also follows a normal distribution (with 0.56\% relative error). The average reduced
$\chi^2$ over 182 runs (two runs are outliers with $\chi^2$ values of 4.8 and 2.9) is 1.03, and the
$\chi^2$ values are distributed with a standard deviation of 0.23 \cite{run107notspecial}. The
anomalous scaling law observed in \cite{kim} is thus not a universal behavior.

\begin{figure}[h!]
\includegraphics[width=13cm]{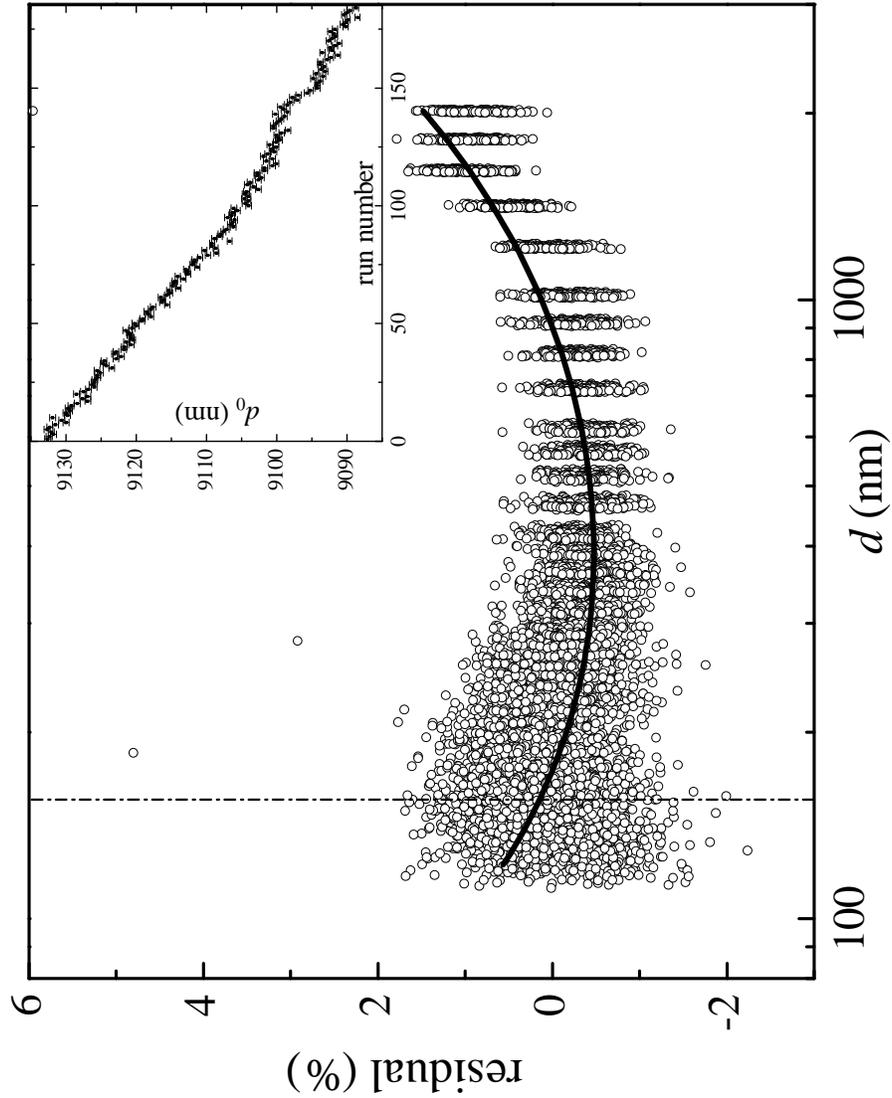}
\caption{Residuals of the fits with $p=1$ plotted as a function of separation. The continuous line
represents the expected deviations due to the use of the proximity force approximation instead of
the whole analytical equation. The dashed line indicates the separation at which the data of Fig.
\ref{alphaerr} were taken. Inset: value of the initial separation between the sphere and the plate
as a function of run number as obtained by fitting the data on the basis of elementary
electrostatic arguments.} \label{inidist}
\end{figure}

Now that the $p=0.7$ behavior is ruled out, it is interesting to plot the residuals of the fits
with $p=1$. From this plot, reported in Fig. \ref{inidist}, it is evident that our data
systematically deviate from the fit, with maximum deviation of $\simeq 1\%$. This behavior is due
to the fact that eq.~\ref{elecforce} is based on the use of the so-called \textit{proximity force
approximation} (PFA) \cite{milonni}. Using the complete analytical equation for the electromagnetic
force between a sphere and a plate \cite{old_book} to calculate the average residual expected from
a fit with eq.~\ref{1overalphaequation} (with $\kappa$ and $d_0$ as free parameters), one obtains
the continuous line of Fig. \ref{inidist}. Note that, although the difference between the complete analytical
equation and the PFA goes to zero as separation decreases, Fig. \ref{inidist} does not show this
behavior. This is due to the fact that we are fitting a linear function (eq. \ref{1overalphaequation}) to slightly curved data.
From the same figure, one can also
see that the random errors at larger separations are actually smaller than that used in the fits
($0.56\%$), which was measured at $d\simeq 150$ nm (see dashed line of Fig. \ref{inidist}). This
justifies the fact that the fit with eq.~\ref{elecforce} and $p=1$ gave $\chi^2 \simeq 1$ even if a
not completely correct theoretical model was used.

We want to stress that such an excellent agreement can only be obtained if the experimental
apparatus is exceptionally stable. In the inset of Fig.~\ref{inidist}, we plot the initial
separation $d_0$ obtained from the fits of all 184~runs. The total drift in $d_0$ is 45~nm over
1050 minutes. This means that our set-up drifts only $\simeq 40$~pm/min (compared to 1~nm/min in
\cite{jourdan_archive}), or 0.24~nm drift per measurement run (compared to $\simeq 60$~nm per run
in \cite{kim}). From the fits, we also observe that the other fit parameter $\kappa$ drifts 1.1\%
over 1050~minutes. This corresponds to 0.001\%/min or 0.006\% per run, and is likely due to a
change of $\gamma$ caused by a slow drift of the laser spot over the photodetector of the AFM head.
The effects of both drifts are negligible in one measurement run.

\begin{figure}[h!]
\includegraphics[width=13cm]{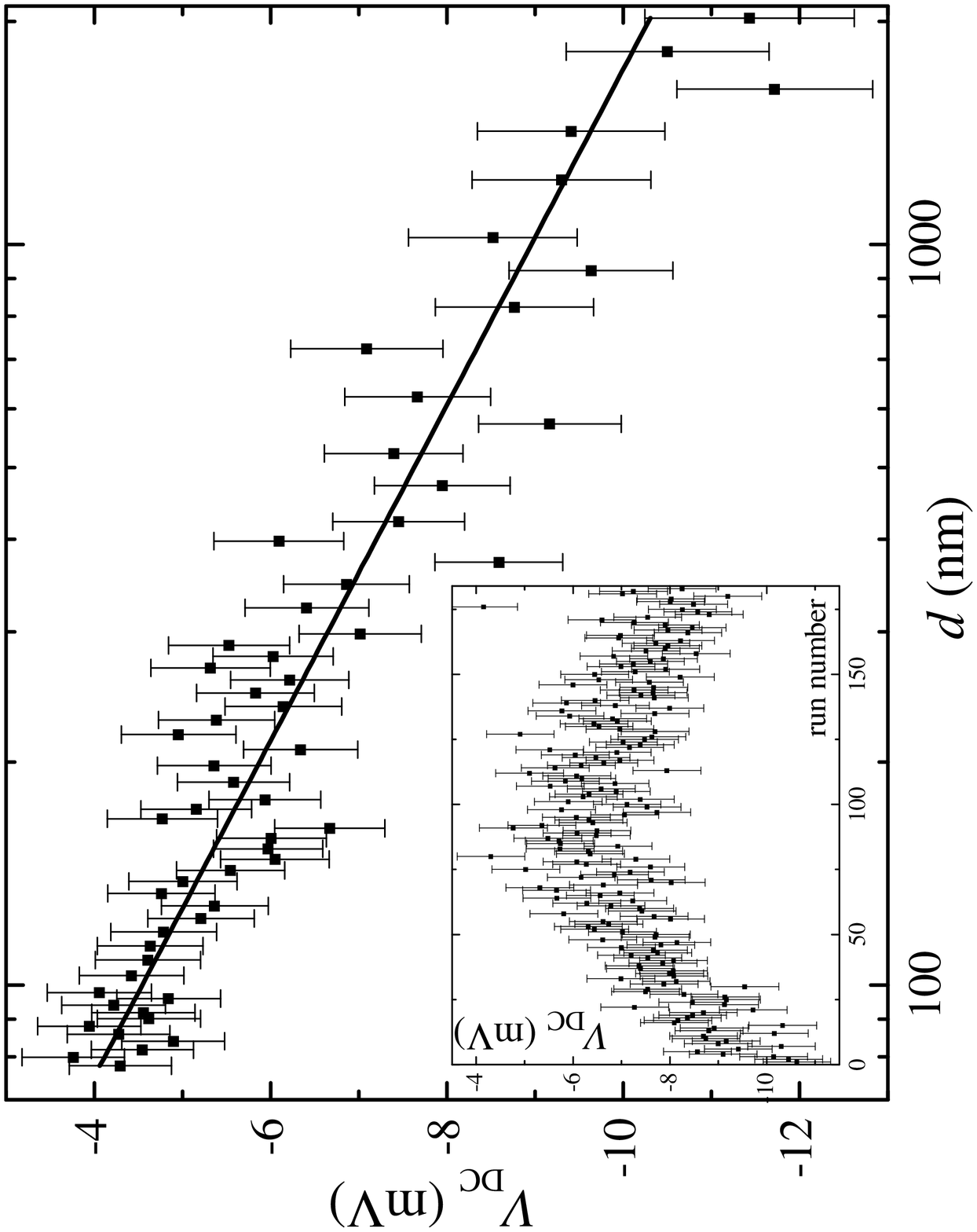}
\centering \caption{Electrostatic compensation voltage as a function of sphere-to-plate separation
for run 107. The continuous line represents the best fit with $V_{DC} = a \log d + b$ (reduced
$\chi^2=0.8$, $a=-4.4 \pm 0.2$~mV, $b=4.3 \pm 0.6$~mV, $d$ in nm). The initial separation between
the sphere and the plate is obtained from the continuous line of Fig.~\ref{run107}. Each error bar
represents the standard deviation of the gaussian distribution of the 184 values of $V_{DC}$ at
that separation. Inset: $V_{DC}$ at $d \simeq 275$ nm as a function of run number.} \label{vdc}
\end{figure}

The authors of \cite{kim} noticed that, in their set-up, the voltage needed to minimize the force
depends on the separation between the surfaces. In Fig.~\ref{vdc} we plot $V_{DC}$ as a function of
$d$ for the data set shown in Fig.~\ref{run107}, where $d_0$ was determined by fitting $\alpha$ as
a function of $d_{pz}$ with $p=1$. Also in our measurements, the compensation voltage clearly
depends on $d$, varying by $\simeq 6$~mV over 2~$\mu$m. This behavior is reproduced in the other
183 measurement runs. In order to show that the dependence observed is not due to drifts of $V_{DC}$ with time, in the inset of Fig.\ref{run107} we plot the value of $V_{DC}$ at $d\simeq 275$ nm as a function of run number. It is clear that, at this separation, $V_{DC}$ drifts, at most, 40 $\mu$V per run. A similar behavior is observed for all separations.

Interestingly, the data seem to distribute along a curve that goes like $a \log d + b$ (reduced $\chi ^2=0.8$). A similar
behavior has been recently reported in an experiment between Ge surfaces \cite{lamoreaux_dv0dd}.
A rigorous explanation of
the origin of the dependence of $V_{DC}$ on $d$ goes beyond the purpose of this paper. Still, our
measurements, together with the results of \cite{lamoreaux_dv0dd} and \cite{kim}, suggest that it
is indeed of fundamental importance to measure $V_{DC}$ for all values of $d$, as in \cite{kim},
and, previously, in \cite{davidepnas2004}. Furthermore, since our calibration follows elementary
electrostatic arguments, we conclude that, in general, it is not sufficient to check the scaling of
$\alpha$ with $d$ to rule out the presence of a distance-dependent $V_{DC}$.

The authors thank F. Mul, J.H. Rector, R.J. Wijngaarden, and R. Griessen for useful discussions.
This work was supported by the Netherlands Organisation for Scientific Research (NWO), under the
Innovational Research Incentives Scheme VIDI-680-47-209. DI acknowledges financial support from the
European Research Council under the European Community's Seventh Framework Programme
(FP7/2007-2013)/ERC grant agreement number 201739.

\pagebreak

\begin{thebibliography}{99}
\bibitem{decca}R.~S.~Decca, D.~L\'{o}pez, E.~Fischbach, G.~L.~Klimchitskaya, D.~E.~Krause, and V.~M.~Mostepanenko, Phys.~Rev.~D~\textbf{75}, 077101 (2007)
\bibitem{kim}W.~J.~Kim, M.~Brown-Hayes, D.~A.~R.~Dalvit, J.~H.~Brownell, and R.~Onofrio, Phys.~Rev.~A~\textbf{78}, 020101(R) (2008)
\bibitem{fluids}This argument does not apply to Casimir force experiments between surfaces in liquids (see J.~N.~Munday and F.~Capasso, Phys.~Rev.~A~\textbf{75}, 060102(R) (2007)),
which, however, have not been used to set new limits on Yukawa corrections to gravity.
\bibitem{lamoreaux_1997}S.~K.~Lamoreaux, Phys.~Rev.~Lett.~\textbf{78}, 5-8 (1997)
\bibitem{kelvinprobe}M.~Nonnenmacher, M.~P.~O'Boyle, and H.~K.~Wickramasinghe,
Appl.~Phys.~Lett.~\textbf{58}, 2921-2923 (1991)
\bibitem{only8datapoints}To avoid that data at very large separations influence the estimate of $V_{AC}$ at closer
distances, we only use the 8 closest points to estimate the next value of $\alpha$.
\bibitem{determine_gamma}To determine this value, we first independently determined the value of $\gamma$. A description of the
technique used to determine $\gamma$ goes beyond the aim of the paper, and, for the sake of
brevity, is omitted.
\bibitem{delyiannis-friend}This frequency is determined by a Deliyannis-Friend filter in the $\omega$
feedback loop.
\bibitem{noteondrifts}Later in the text, we show that the drift in $\kappa$ is negligible on the scale of Fig. \ref{alphaerr}.
\bibitem{run107notspecial}Note that the value of the reduced $\chi^2$ for run~107 is actually
higher than the average $\chi^2$ of all runs. The data presented in Fig.~\ref{run107} are thus
certainly not among the ones that better match the theory.
\bibitem{milonni} P. W. Milonni, The quantum vacuum: an introduction to quantum electrodynamics (Academic Press, San Diego, 1993)
\bibitem{old_book} W. R. Smythe, Static and dynamic electricity (McGraw Hill, New York, 1939)
\bibitem{jourdan_archive}G.~Jourdan, A.~Lambrecht, F.~Comin, and J.~Chevrier, e-print
ArXiv:0712.1767
\bibitem{lamoreaux_dv0dd}S.~K.~Lamoreaux, e-print ArXiv:0808.0885 (2008)
\bibitem{davidepnas2004}D.~Iannuzzi, M.~Lisanti, and F.~Capasso, Proc. Nat. Ac. Sci. USA~\textbf{101}, 4019-4023
(2004)





\end{thebibliography}
\end{document}